\documentclass[11pt,a4paper]{article}
\usepackage{acl2015}
\usepackage{times}
\usepackage{url}
\usepackage{latexsym}
\usepackage{multirow}
\usepackage{graphicx}
\usepackage{times}
\usepackage{latexsym}
\usepackage{amsmath}
\usepackage{amssymb}
\usepackage{paralist}
\usepackage{url}
\usepackage{paralist}
\usepackage{booktabs}
\usepackage[caption=false]{subfig}

\usepackage{algorithm}
\usepackage{algorithmic}

\usepackage{tikz}

\newcommand\idea[1]{\textcolor{olive}{}} 

\newcommand\Mark[1]{\textsuperscript#1}

\title{Classifying Tweet Level Judgements of Rumours in Social Media}

\author{%
Michal Lukasik,\Mark{1} Trevor Cohn\Mark{2} \and Kalina Bontcheva\Mark{1} \\
\hspace{-5ex}\begin{tabular}{*{2}{c}}
\Mark{1}Computer Science & \hspace{-4ex}\Mark{2}Computing and Information Systems \tabularnewline
The University of Sheffield & The University of Melbourne \tabularnewline
\texttt{\{m.lukasik,k.bontcheva\}@shef.ac.uk} & \texttt{t.cohn@unimelb.edu.au}
\end{tabular}
}

\date{}

\begin{document}
\maketitle
\begin{abstract}
Social media is a rich source of rumours and corresponding community reactions. 
Rumours reflect different characteristics, some shared and some individual. 
We formulate the problem of classifying tweet level judgements of rumours as a supervised learning task. 
Both supervised and unsupervised domain adaptation are considered, in which tweets from a rumour are classified on the basis of other annotated rumours.
We demonstrate how multi-task learning helps achieve good results on rumours from the 2011 England riots.
\end{abstract}

\section{Introduction}

There is an increasing need to interpret and act upon rumours spreading quickly through social media, especially in circumstances  where their veracity is hard to establish. For instance, during an earthquake in Chile rumours  spread through Twitter that a volcano had become active and that there was a tsunami warning in Valparaiso \cite{Mendoza-rumours-2010}. Other examples, from the riots in 
England in 2011, were that rioters were going to attack Birmingham's children hospital and 
that animals had escaped from the zoo \cite{procter2013reading}.

Social scientists \cite{procter2013reading} analysed manually a sample of tweets expressing different judgements towards rumours and categorised them manually in supporting, denying or questioning.
The goal here is to carry out tweet-level judgement classification automatically, in order to assist in (near) real-time rumour monitoring by journalists and authorities \cite{procter2013reading}. In addition, information about tweet-level judgements has been used as a first step for early rumour detection by \cite{zhao2015www}.

\begin{table}
\resizebox{\columnwidth}{!}{
\begin{tabular}{p{0.75\linewidth}l}
\toprule text & position\\
\midrule
Birmingham Children's hospital has been attacked. F***ing morons. \#UKRiots & support \\
\midrule
Girlfriend has just called 
her ward in Birmingham Children's
Hospital \& there's no sign 
of any trouble \#Birminghamriots & deny \\
\midrule
Birmingham children's hospital guarded by police? Really? Who  would target a childrens hospital \#disgusting \#Birminghamriots & question \\
\bottomrule
\end{tabular}}
\caption{Tweets on a rumour about hospital being attacked during 2011 England Riots.}
\label{tab:tweets_hospital}
\end{table}



The focus here is on tweet-level judgement classification on unseen rumours, based on a training set of other already annotated rumours. 
Previous work on this problem either considered unrealistic settings ignoring temporal ordering and rumour identities \cite{Qazvinian:2011:RIM:2145432.2145602} or proposed regular expressions as a solution \cite{zhao2015www}. 
We expect  posts expressing similar opinions to exhibit many similar characteristics across different rumours. 
Based on the assumption of a common underlying linguistic signal, we build a transfer learning system that labels newly emerging rumours for which we have little or no annotated data. 
Results demonstrate that Gaussian Process-based multi task learning allows for significantly improved performance.


The novel contributions of this paper are:
\begin{inparaenum}
\item Formulating the problem of classifying judgements of rumours in both supervised and unsupervised domain adaptation settings.
\item Showing how a multi-task learning approach outperforms single-task methods.
\end{inparaenum}

\section{Related work}
\label{sec:related_work}
In the context of rumour spread in social media, researchers have studied differences in information flows between content of varying credibility. 
For instance, \newcite{procter2013reading} grouped source tweets and re-tweets into ‘information flows’ \cite{Lotan-flows-2011}, 
then ranked these by flow size, as a proxy of significance. 
Information flows were then categorised manually. 
Along similar vein, 
\newcite{Mendoza-rumours-2010} found that users deal with ‘true’ and ‘false’ rumours differently: the 
former are affirmed more than 90\% of the time, whereas the latter are challenged (questioned or denied) 50\% 
of the time. 
\newcite{ICWSM148122} analyzed a set rumours  from the \emph{Snopes.com} website that have been matched to Facebook public conversations. They concluded that false rumours are more likely to receive a comment with link to \emph{Snopes.com} website. However, none of the above attempted to automatically classify rumours.

With respect to automatic methods for detecting misinformation and disinformation in social media, \newcite{Ratkiewicz-abuse-2011} 
detect political abuse (a kind of disinformation) spread through Twitter.
The task is defined in purely information diffusion settings and is not necessarily related with the truthfulness of the piece of information. 
\newcite{Mendoza-credibility-2013} proposed methods for identifying newsworthy information 
cascades on Twitter and then classifying these cascades as credible and not credible. 
The main difference from our task is 
that credibility classification is carried out over the entire information cascade, classified objects are not necessarily rumours and no explicit judgement classification was performed in their approach. 

Early rumour identification is the focus of \newcite{zhao2015www}, where regular expressions are used for finding questioning and denying tweets as a key pre-requisite step for rumour detection.
Unfortunately, when we applied these regular expressions on our dataset, they yielded only 16\% recall for questioning and 14\% recall for denying tweets. 
Consequently, this motivated us to seek a better approach to tweet-level classification.  

The work most relevant to ours is due to~\newcite{Qazvinian:2011:RIM:2145432.2145602}. Their method first carries out rumour retrieval, whereby tweets are classified into rumour related and non-rumour related. Next, rumour-related tweets are classified into supporting and not-supporting. The classifier is trained by ignoring rumour identities, i.e., pooling together tweets from all rumours, and ignoring the temporal dependencies between tweets.
In contrast, we formulate the rumour classification problem as transfer learning, where unseen rumours (or rumours with few initial tweets observed) are classified using already known rumours -- a much harder and more practical setting. 
Moreover, unlike \newcite{Qazvinian:2011:RIM:2145432.2145602}, we consider the multi-class classification problem and do not collaps questioning and denying tweets into a single class, since they differ significantly. 

\idea{More related work, more papers.}

\section{Data}
\label{sec:data}
\newcommand\no{$\times$}
\begin{table}
\centering
\footnotesize
\begin{tabular}{lrrr}
\toprule 
Rumour & Supporting & Denying & Questioning\\
\midrule 
army bank & 62 & 42 & 73\\
hospital & 796 & 487 & 132\\
London Eye & 177 & 295 & 160\\
McDonald's & 177 & 0 & 13\\
Miss Selfridge's & 3150 & 0 & 7\\
police beat girl & 783 & 4 & 95\\
zoo & 616 & 129 & 99\\
\bottomrule 
\end{tabular}}
\caption{Counts of tweets with supporting, denying or questioning labels in each rumour collection.}{
\label{table:Datacounts}
\end{table}


We evaluate our work on several rumours circulating on Twitter during the England riots in 2011 (see Table~\ref{table:Datacounts}). The dataset was analysed and annotated manually as supporting, questioning, or denying a rumour, by a team of social scientists studying the role of social media during the riots \cite{procter2013reading}. 
%
%
The original dataset also included commenting tweets, but these have been removed from our experiments due to their small number (they constituted only 5\% of the corpus).


As can be seen from the dataset overview in Table~\ref{table:Datacounts}, different rumours exhibit varying proportions of supporting, denying and questioning tweets, which was also observed in other studies of rumours \cite{Mendoza-rumours-2010,Qazvinian:2011:RIM:2145432.2145602}.
These variations in majority classes across rumours underscores the modeling challenge in tweet-level classification of rumour attitudes.

With respect to veracity, one rumour has been confirmed as true (Miss Selfridge's being on fire), one is unsubstantiated (police beat girl), and the remaining five are known to be false. 
Note, however, that the focus here is not on classifying truthfulness, but instead on identifying the attitude expressed in each tweet towards the rumour. 

\idea{Refer to Figure~\ref{fig:temporal_points}}



\section{Problem formulation}
\label{sec:problem}
Let $R$ be a set of rumours, each of which consists of tweets discussing it, $\forall_{r \in R}$ $T_r$ $= \{t^r_1, \cdots, t^r_{r_n}\}$. $T = \cup_{r \in R} T_r$ is the complete set of tweets from all rumours. Each tweet is classified as supporting, denying or questioning with respect to its rumour: $y(t) \in \{0, 1, 2\}$, where $0$ denotes supporting, $1$ means denying and $2$ denotes questioning. 

First, we consider the Leave One Out (LOO) setting, which means that for each rumour $r \in R$, we construct the test set equal to $T_r$ and the training set equal to $T \setminus T_r$. Therefore this is a very challenging and realistic scenario, where the test set contains an entirely unseen rumour, from those in the training set.

The second setting is Leave Part Out (LPO). In this formulation, a very small number of initial tweets from the target rumour is added to the training set $\{t^r_1, \cdots, t^r_{{{r_k}}}\}$.
This scenario becomes applicable typically soon after a rumour breaks out and journalists have started monitoring and analysing the related tweet stream.
The experiments section investigates how the number of initial training tweets influences classification performance on a fixed test set, namely: $\{t^r_{{{r_l}}{}}, \cdots, t^r_{r_n}\}$, $l>k$.

The tweet-level classification problem here assumes that tweets from the training set are already labelled with the rumour discussed and the attitude expressed towards that. 
This information can be acquired either via manual annotation as part of expert analysis, as is the case with our dataset, or automatically, e.g. using pattern-based rumour detection \cite{zhao2015www}. 
Afterwards, our method can be used to classify the attitudes expressed in each new tweet from outside the training set. 

\section{Gaussian Processes for Classification}
\label{sec:methods_description}
\label{sec:methods}

%

Gaussian Processes are a Bayesian non-parametric machine learning framework that has been shown to work well for a range of NLP problems, often beating other state-of-the-art methods \cite{Cohn13modellingannotator,DBLP:conf/eacl/LamposAPC14,beck-etal_EMNLP:2014,jobs15acl}. 
We use Gaussian Processes as this probabilistic kernelised framework avoids the need for expensive cross-validation for hyperparameter selection.\footnote{There exist frequentist kernel methods, like SVMs, which additionally require extensive heldout parameter tuning.} 

The central concept of Gaussian Process Classification (GPC; \cite{Rasmussen:2005:GPM:1162254}) is a latent function $f$ over inputs \mbox{$\mathbf{x}$: $f(\mathbf{x}) \sim\ \mathcal{GP}(m(\mathbf{x}), k(\mathbf{x}, \mathbf{x}'))$}, where $m$ is the mean function, assumed to be $0$ and $k$ is the kernel function, specifying the degree to which the outputs covary as a function of the inputs. We use a 
linear kernel,
$k(\mathbf{x}, \mathbf{x}') = \sigma^2 \mathbf{x}^{\top}\mathbf{x}'$.
The latent function is then mapped by the probit function $\Phi(f)$ into the range $[0, 1]$, such that the resulting value can be interpreted as $p(y=1 | \mathbf{x})$.

The GPC posterior is calculated as
\vspace{-2mm}
\begin{equation*}
p(f^* | X, \mathbf{y}, \mathbf{x_*}) = \int p(f^* | X, \mathbf{x_*}, \mathbf{f}) \frac{p(\mathbf{y} | \mathbf{f})p(\mathbf{f})}{p(\mathbf{y}|X)} d\mathbf{f} \, \!,
\end{equation*}
where  $p(\mathbf{y}|\mathbf{f}) = \displaystyle \prod_{j=1}^{n} \Phi(f_j)^{y_j} (1 - \Phi(f_j))^{1-y_j}$ is the Bernoulli likelihood of class $y$. After calculating the above posterior from the training data, this is used in prediction, i.e.,
\begin{equation*}
p(y_* \!=\! 1|X, \mathbf{y}, \mathbf{x_*}) \!=\!\!
\int \Phi\left(f_*\right)p\left(f_*|X, \mathbf{y}, \mathbf{x_*}\right)df_* \, .
\vspace{-2mm}
\end{equation*}

The above integrals are intractable and approximation techniques are required to solve them. 
There exist various methods to deal with calculating the posterior; here we use Expectation Propagation (EP; \cite{Minka:2002:EGA:2073876.2073918}). 
In EP, the posterior is approximated by a fully factorised distribution, where each component is assumed to be an unnormalised Gaussian.

In order to conduct multi-class classification, we perform a one-vs-all classification for each label and then assign the one with the highest likelihood, amongst the three (supporting, denying, questioning).
We choose this method due to interpretability of results, similar to recent work on occupational class classification \cite{jobs15acl}.

\idea{More on GPs}

\vspace{-5mm}
\paragraph{Intrinsic Coregionalization Model}
In the LPO setting initial labelled tweets from the target rumour are observed as well.
In this case, we propose to weight the importance of tweets from the reference rumours depending on how similar their characteristics are to the tweets from the target rumour available for training.
To handle this with  GPC, we use a multiple output model based on the Intrinsic Coregionalisation Model (ICM; \cite{Alvarez:2012:KVF:2344402.2344403}). 
It has already been applied successfully to NLP regression problems \cite{beck-etal_EMNLP:2014} and it can also be applied to classification ones. 
ICM parametrizes the kernel by a matrix which represents the extent of covariance between pairs of tasks. 
The complete kernel takes form of
\begin{equation*}
k((\mathbf{x}, d), (\mathbf{x}', d')) = k_{data}(\mathbf{x}, \mathbf{x}') B_{d, d'} \, ,
\end{equation*}
where B is a square coregionalization matrix, $d$ and $d'$ denote the tasks of the two inputs and $k_{data}$ is a kernel for comparing inputs $\mathbf{x}$ and $\mathbf{x}'$ (here, linear).
We parametrize the coregionalization matrix \mbox{$B=\boldsymbol{\kappa} I+\boldsymbol{vv}^T$}, where $\boldsymbol{v}$ specifies the correlation between tasks and the vector $\mathbf{\kappa}$ controls extent of task independence. 

\paragraph{Hyperparameter selection}
We tune hyperparameters $\mathbf{v}$, $\boldsymbol{\kappa}$ and $\sigma^2$ by maximizing evidence of the model $p(\mathbf{y}|X)$, thus having no need for a validation set.

\paragraph{Methods}
We consider GPs in three different settings, varying in what data the model is trained on and what kernel it uses. 
The first setting (denoted GP) considers only target rumour data for training. 
The second (GPPooled) additionally considers tweets from reference rumours (i.e. other than the target rumour).
The third setting is GPICM, where an ICM kernel is used to weight influence from tweets from reference rumours.

\section{Features}

We conducted a series of preprocessing steps in order to address data sparsity. 
All words were lowercased; stopwords removed; all emoticons were replaced with words\footnote{We used the dictionary from:
 \url{http://bit.ly/1rX1Hdk} and extended 
 it with: :o, $:|$, =/, :s, :S, :p.}; and stemming was performed. 
 In addition, multiple occurrences of a character were replaced with a double occurrence \cite{Agarwal:2011:SAT:2021109.2021114}, to correct for misspellings and lengthenings, e.g., \emph{looool}. 
 All punctuation was also removed, except for \emph{.}, \emph{!} and \emph{?}, which we hypothesize to be important for expressing emotion.
 Lastly, usernames were removed as they tend to be rumour-specific, i.e., very few users comment on more than one rumour.

After preprocessing the text data, we use either the resulting bag of words (BOW) feature representation or replace all words with their Brown
cluster ids (Brown), using 1000 clusters acquired from a large scale Twitter corpus \cite{Owoputi13improvedpart-of-speech}.
In all cases, simple re-tweets are removed from the training set to prevent bias \cite{LLEWELLYN14.845}.

\section{Experiments and Discussion}
\label{sec:experiments}

Table~\ref{tab:loro_notarged} shows the mean accuracy in the LOO scenario following the GPPooled method, which pools all reference rumours together ignoring their task identities. ICM can not use correlations to target rumour in this case and so can not be used. 
The majority baseline simply assigns the most frequent class from the training set.

We can observe that methods perform on a level similar to majority vote, outperforming it only slightly.
This indicates how difficult the LOO task is, when no annotated target rumour tweets are available.

\begin{table}
\begin{center}
{
\begin{tabular}{ll}
\toprule 
method & acc\\
\midrule 
Majority & 0.68\\
GPPooled Brown & 0.72\\
GPPooled BOW & 0.69\\
\bottomrule 
\end{tabular}}
\end{center}
\caption{Accuracy taken across all rumours in the LOO setting.}
\label{tab:loro_notarged}
\end{table}

\begin{figure*}
    \centering
    \includegraphics[width=1\textwidth]{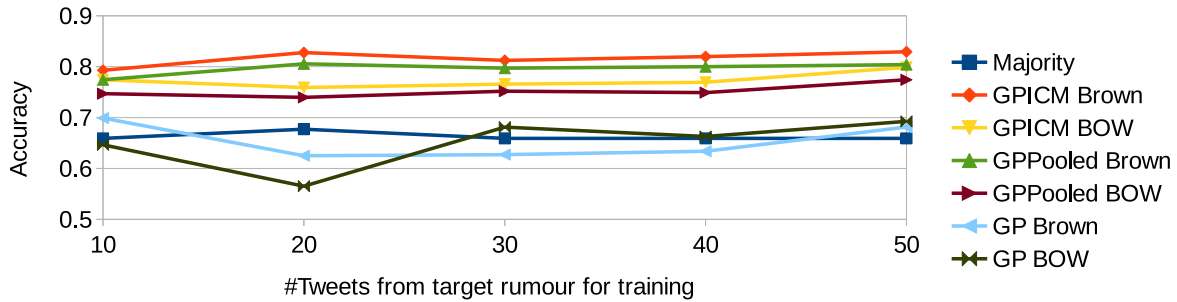}
    \caption{Accuracy measures for different methods versus the size of the target rumour used for training in the LPO setting. The test set is fixed to all but the first 50 tweets of the target rumour.}
    \label{fig:increase_training_rumour}
\end{figure*}

Figure~\ref{fig:increase_training_rumour} shows accuracy for a range of methods as the number of tweets about the target rumour used for training increases. 
Most notably, performance increases from 70\% to around 80\%, after only 10 annotated tweets from the target rumour become available, as compared to the results on unseen rumours from Table~\ref{tab:loro_notarged}. However, as the amount of target rumour increases, performance does not increase further, which suggests that even only 10 human-annotated tweets are enough to achieve significant performance benefits. 
Note also how the use of reference rumours is very important, as methods using only the target rumour obtain accuracy similar to the Majority vote classifier (GP Brown and GP BOW).

\begin{table}
{
\begin{tabular}{lrr}
\toprule 
supporting & denying & questioning\\
\midrule 
? & fake & ? \\
{\small 10001101 } & {\small 11111000001 } & {\small 10001101 } \\ \midrule
! & not & ! \\ 
{\small 10001100 } & {\small 001000 } & {\small 10001100 } \\ \midrule
not & ? & hope \\
{\small 001000 } & {\small 10001101 } & {\small 01000111110 } \\ \midrule
fake & ! & true \\ 
{\small 11111000001 } & {\small 10001100 } & {\small 111110010110 } \\ \midrule
true & bullshit & searching \\
{\small 111110010110 } & {\small 11110101011111 } & {\small 01111000010 } \\
\bottomrule 
\end{tabular}}
\caption{Top 5 Brown clusters, each shown with a representative
  word. For further details please see the cluster definitions at \url{http://www.ark.cs.cmu.edu/TweetNLP/cluster_viewer.html}.}
\label{tab:weights_army_bank}
\end{table}

The top performing methods are GPCIM and GPPooled, where use of Brown clusters consistently improves results for both methods over BOW, irrespective of the number of tweets about the target rumour annotated for training. 
Moreover, GPICM is better than GPPooled both with Brown and BOW features and GPCIM with Brown is ultimately the best performing of all. 

In order to analyse the importance of Brown clusters, Automatic Relevance Determination (ARD) is used \cite{Rasmussen:2005:GPM:1162254} for the best performing GPICM Brown in the LPO scenario. 
Only the case where the first 10 tweets are used for training is considered, since it already performs very well.
Using ARD, we learn a separate length-scale for each feature, thus establishing their importance.
The weights learnt for different clusters are averaged over the 7 rumours and the top 5 Brown clusters for each label are shown in Table~\ref{tab:weights_army_bank}.
We can see that clusters around the words \emph{fake} and \emph{bullshit} turn out to be important for the denying class,
 and \emph{true} for both supporting and questioning classes. 
This reinforces our hypothesis that common linguistic cues can be found across multiple rumours. 
Note how punctuation proves important as well, since clusters \emph{?} and \emph{!} are also very prominent. 

\section{Conclusions}
This paper investigated the problem of classifying judgements expressed in tweets about rumours.
First, we considered a setting where no training data from target rumour is available (LOO).
Without access to annotated examples of the target rumour the
learning problem becomes very difficult.  
We showed that in the supervised domain adaptation setting  (LPO) even annotating a small number of tweets helps  to  achieve better results.
Moreover, we demonstrated the benefits of a multi task learning approach, as well as that Brown cluster features are more useful for the task than simple bag of words.

Judgement estimation is undoubtedly of great value e.g. for marketing, politics and journalism, helping to target widely believed topics. Although the focus here is on classifying community reactions, \newcite{Mendoza-credibility-2013} showed that community reaction is correlated with actual rumour veracity. 
Consequently our classification methods may prove useful in the broader and more challenging task of annotating veracity.

An interesting direction for future work would be adding non-textual features. 
For example, the rumour diffusion pattern \cite{lukasik15} may be a useful cue for judgement classification.

\section*{Acknowledgments}
Work partially supported by the European Union under grant agreement No. 611233 {\sc Pheme}.
The work was implemented using the GPy toolkit \cite{gpy2014}.


\bibliographystyle{acl}
\bibliography{bibliography}

\begin{thebibliography}{}

\bibitem[\protect\citename{Agarwal \bgroup et al.\egroup
  }2011]{Agarwal:2011:SAT:2021109.2021114}
Apoorv Agarwal, Boyi Xie, Ilia Vovsha, Owen Rambow, and Rebecca Passonneau.
\newblock 2011.
\newblock Sentiment analysis of twitter data.
\newblock In {\em Proceedings of the Workshop on Languages in Social Media},
  LSM '11, pages 30--38.

\bibitem[\protect\citename{\'{A}lvarez \bgroup et al.\egroup
  }2012]{Alvarez:2012:KVF:2344402.2344403}
Mauricio~A. \'{A}lvarez, Lorenzo Rosasco, and Neil~D. Lawrence.
\newblock 2012.
\newblock Kernels for vector-valued functions: A review.
\newblock {\em Found. Trends Mach. Learn.}, 4(3):195--266.

\bibitem[\protect\citename{Beck \bgroup et al.\egroup
  }2014]{beck-etal_EMNLP:2014}
Daniel Beck, Trevor Cohn, and Lucia Specia.
\newblock 2014.
\newblock Joint emotion analysis via multi-task {Gaussian} processes.
\newblock In {\em Proceedings of the Conference on Empirical Methods in Natural
  Language Processing}, EMNLP '14, pages 1798--1803.

\bibitem[\protect\citename{Castillo \bgroup et al.\egroup
  }2013]{Mendoza-credibility-2013}
Carlos Castillo, Marcelo Mendoza, and Barbara Poblete.
\newblock 2013.
\newblock Predicting information credibility in time-sensitive social media.
\newblock {\em Internet Research}, 23(5):560--588.

\bibitem[\protect\citename{Cohn and Specia}2013]{Cohn13modellingannotator}
Trevor Cohn and Lucia Specia.
\newblock 2013.
\newblock Modelling annotator bias with multi-task {Gaussian} processes: An
  application to machine translation quality estimation.
\newblock In {\em 51st Annual Meeting of the Association for Computational
  Linguistics}, ACL '13, pages 32--42.

\bibitem[\protect\citename{Friggeri \bgroup et al.\egroup }2014]{ICWSM148122}
Adrien Friggeri, Lada Adamic, Dean Eckles, and Justin Cheng.
\newblock 2014.
\newblock Rumor cascades.
\newblock In {\em International AAAI Conference on Weblogs and Social Media}.

\bibitem[\protect\citename{{GPy}{\ }authors}2015]{gpy2014}
The {GPy}{\ }authors.
\newblock 2015.
\newblock {GPy: A Gaussian process framework in Python}.
\newblock \url{http://github.com/SheffieldML/GPy}.

\bibitem[\protect\citename{Lampos \bgroup et al.\egroup
  }2014]{DBLP:conf/eacl/LamposAPC14}
Vasileios Lampos, Nikolaos Aletras, Daniel Preotiuc{-}Pietro, and Trevor Cohn.
\newblock 2014.
\newblock Predicting and characterising user impact on twitter.
\newblock In {\em Proceedings of the 14th Conference of the European Chapter of
  the Association for Computational Linguistics}, EACL'14, pages 405--413.

\bibitem[\protect\citename{Llewellyn \bgroup et al.\egroup
  }2014]{LLEWELLYN14.845}
Clare Llewellyn, Claire Grover, Jon Oberlander, and Ewan Klein.
\newblock 2014.
\newblock Re-using an argument corpus to aid in the curation of social media
  collections.
\newblock In {\em Proceedings of the Ninth International Conference on Language
  Resources and Evaluation}, LREC'14, pages 462--468.

\bibitem[\protect\citename{Lotan \bgroup et al.\egroup }2011]{Lotan-flows-2011}
Gilad Lotan, Erhardt Graeff, Mike Ananny, Devin Gaffney, Ian Pearce, and danah
  boyd.
\newblock 2011.
\newblock The {Arab} spring| the revolutions were tweeted: Information flows
  during the 2011 {Tunisian} and {Egyptian} revolutions.
\newblock {\em International Journal of Communication}, 5(0).

\bibitem[\protect\citename{Lukasik \bgroup et al.\egroup }2015]{lukasik15}
Michal Lukasik, Trevor Cohn, and Kalina Bontcheva.
\newblock 2015.
\newblock Point process modelling of rumour dynamics in social media.
\newblock In {\em Proceedings of the 53rd Annual Meeting of the Association for
  Computational Linguistics and the 7th International Joint Conference on
  Natural Language Processing of the Asian Federation of Natural Language
  Processing, {ACL} 2015}, pages 518--523.

\bibitem[\protect\citename{Mendoza \bgroup et al.\egroup
  }2010]{Mendoza-rumours-2010}
Marcelo Mendoza, Barbara Poblete, and Carlos Castillo.
\newblock 2010.
\newblock Twitter under crisis: Can we trust what we {RT}?
\newblock In {\em 1st Workshop on Social Media Analytics}, SOMA'10, pages
  71--79.

\bibitem[\protect\citename{Minka and
  Lafferty}2002]{Minka:2002:EGA:2073876.2073918}
Thomas Minka and John Lafferty.
\newblock 2002.
\newblock Expectation-propagation for the generative aspect model.
\newblock In {\em Proceedings of the Eighteenth Conference on Uncertainty in
  Artificial Intelligence}, UAI'02, pages 352--359.

\bibitem[\protect\citename{Owoputi \bgroup et al.\egroup
  }2013]{Owoputi13improvedpart-of-speech}
Olutobi Owoputi, Chris Dyer, Kevin Gimpel, Nathan Schneider, and Noah~A. Smith.
\newblock 2013.
\newblock Improved part-of-speech tagging for online conversational text with
  word clusters.
\newblock In {\em Proceedings of NAACL}, pages 380--390.

\bibitem[\protect\citename{Preotiuc{-}Pietro \bgroup et al.\egroup
  }2015]{jobs15acl}
Daniel Preotiuc{-}Pietro, Vasileios Lampos, and Nikolaos Aletras.
\newblock 2015.
\newblock An analysis of the user occupational class through twitter content.
\newblock In {\em Proceedings of the 53rd Annual Meeting of the Association for
  Computational Linguistics and the 7th International Joint Conference on
  Natural Language Processing of the Asian Federation of Natural Language
  Processing, {ACL} 2015}, pages 1754--1764.

\bibitem[\protect\citename{Procter \bgroup et al.\egroup
  }2013]{procter2013reading}
Rob Procter, Jeremy Crump, Susanne Karstedt, Alex Voss, and Marta Cantijoch.
\newblock 2013.
\newblock Reading the riots: What were the police doing on twitter?
\newblock {\em Policing and society}, 23(4):413--436.

\bibitem[\protect\citename{Qazvinian \bgroup et al.\egroup
  }2011]{Qazvinian:2011:RIM:2145432.2145602}
Vahed Qazvinian, Emily Rosengren, Dragomir~R. Radev, and Qiaozhu Mei.
\newblock 2011.
\newblock Rumor has it: Identifying misinformation in microblogs.
\newblock In {\em Proceedings of the Conference on Empirical Methods in Natural
  Language Processing}, EMNLP '11, pages 1589--1599.

\bibitem[\protect\citename{Rasmussen and
  Williams}2005]{Rasmussen:2005:GPM:1162254}
Carl~Edward Rasmussen and Christopher K.~I. Williams.
\newblock 2005.
\newblock {\em Gaussian Processes for Machine Learning (Adaptive Computation
  and Machine Learning)}.
\newblock The MIT Press.

\bibitem[\protect\citename{Ratkiewicz \bgroup et al.\egroup
  }2011]{Ratkiewicz-abuse-2011}
Jacob Ratkiewicz, Michael Conover, Mark Meiss, Bruno Gonçalves, Alessandro
  Flammini, and Filippo Menczer.
\newblock 2011.
\newblock Detecting and tracking political abuse in social media.
\newblock In {\em 5th International AAAI Conference on Weblogs and Social
  Media}, ICWSM'11.

\bibitem[\protect\citename{Zhao \bgroup et al.\egroup }2015]{zhao2015www}
Zhe Zhao, Paul Resnick, and Qiaozhu Mei.
\newblock 2015.
\newblock Early detection of rumors in social media from enquiry posts.
\newblock In {\em International World Wide Web Conference Committee (IW3C2)}.

\end{thebibliography}
\end{document}